\documentclass{article}
\usepackage{ijcai19}

\usepackage{times}
\usepackage{soul}
\usepackage{url}
\usepackage[hidelinks]{hyperref}
\usepackage[utf8]{inputenc}
\usepackage[small]{caption}
\usepackage{graphicx}
\usepackage{amsmath}
\usepackage{booktabs}
\usepackage{multirow}
\usepackage{color}
\usepackage{enumitem}

\title{Semantic Segmentation of Seismic Images}

\author{
Daniel Civitarese$^1$\footnote{Contact Author}\and
Daniela Szwarcman$^{1, 2}$\and
Emilio Vital Brazil$^1$\And
Bianca Zadrozny$^1$\\
\affiliations
$^1$IBM Research\\
$^2$PUC-Rio\\
\emails
\{sallesd, daniszw, evital, biancaz\}@br.ibm.com
}

\begin{document}

\maketitle

\begin{abstract}
Almost all work to understand Earth’s subsurface on a large scale relies on the interpretation of seismic surveys by experts who segment the survey (usually a cube) into layers; a process that is very time demanding.
In this paper, we present a new deep neural network architecture specially designed to semantically segment seismic images with a minimal amount of training data. To achieve this, we make use of a transposed residual unit that replaces the traditional dilated convolution for the decode block. Also, instead of using a predefined shape for up-scaling, our network learns all the steps to upscale the features from the encoder.
We train our neural network using the Penobscot 3D dataset; a real seismic dataset acquired offshore Nova Scotia, Canada.
We compare our approach with two well-known deep neural network topologies: Fully Convolutional Network and U-Net.
In our experiments, we show that our approach can achieve more than 99\% of the mean intersection over union (mIOU) metric, outperforming the existing topologies. Moreover, our qualitative results show that the obtained model can produce masks very close to human interpretation with very little discontinuity.
\end{abstract}

\section{Introduction}

\noindent
Seismic surveys are the most used data to understand the subsurface nowadays, which is an indirect measure that has a long and complex chain of computational processing.

In one of the crucial steps in seismic interpretation, the expert must find the separation between two bodies of rock having different physical properties; this separation (the contact between two strata) can be called horizons. 
To be able to mark the horizons the geoscientist analyzes slices (images) of the seismic cube (survey) looking for visual patterns that could indicate the strata \cite{randen2000three}.  
This task is time-consuming and it overloads human interpreters as the amount of geophysical information is continually increasing.
Typically the expert must analyze more than 20\% of all lines of the seismic cube, which can mean thousands of images in some recent surveys.

Both industry and academia have been advancing the performance of automated and semi-automated algorithms to extract features from seismic images \cite{randen2000three}.
In the past, most of the previous attempts to automate parts of the seismic facies analysis process relied mainly on traditional computer vision (CV) techniques.
For instance, \cite{gao2011latest} studied the application of gray level co-occurrence matrix (GLCM) methods;
some more recent work introduced other imaging processing techniques like phase congruency \cite{Shafiq2017}, the gradient of texture (GoT) \cite{Wang2016}, local binary patterns (LBP) \cite{mattos2017assessing} or even combining more than one method \cite{Ferreira2016}.
In the last two years, the interest in applying machine learning techniques to process seismic data has increased, and last year there was even a workshop at NeurIPS called \emph{Machine Learning for Geophysical \& Geochemical Signals}.

\cite{chevitarese2018deep} present a study about seismic facies classification using convolutional neural networks.
They test the models on two real-world seismic datasets \cite{baroni2018netherlands,baroni2018penobscot} with promising results.
In \cite{chevitarese2018transfer}, the authors show how transfer learning may be used to accelerate and improve training on seismic images, and in \cite{chevitarese2018seismic} they applied the transfer technique to semantically segment rock layers.

However, all these articles lack a detailed discussion about the effect of different network topologies, training procedures, and parameter choices, being focused on the domain application rather than on the machine learning aspects.
To address this lack of detail, \cite{chevitarese2018efficient} present deep learning models specifically for the task of classification of seismic facies along with a detailed discussion on how different parameters may affect the model's performance.
In this work, they present four different topologies: Danet1, Danet2, Danet3, and Danet4, that are able to reach 90\% of classification accuracy with less than 1\% of the available data.

\subsection{Semantic segmentation of seismic images}
Seismic facies segmentation consists of generating dense pixel-wise predictions that delineates the horizons in seismic images. To the best of our knowledge, \cite{chevitarese2018seismic} presented the first study on applying deep learning models to seismic facies segmentation.
The authors in \cite{alaudah2019machine} presented a similar work using the same dataset based on deconvolutional neural networks, but showing not so good results in the final segmentation.

Additionally to the stratigraphic interpretation of an entire seismic cube, other works have demonstrated how CNN can successfully segment fragments of seismic data. 
In \cite{waldeland2018convolutional,shi2018automatic,zeng2018automatic}, a CNN is used to identify an entire salt structure where the output image is very close to the ground truth.
The authors in \cite{zeng2018automatic,karchevskiy2018automatic} uses a  segmentation process to delineate salt domes in a given seismic.

\cite{chevitarese2018seismic} introduces a deep neural network topology for segmentation showing good results on seismic data based on existing work on fully convolutional networks \cite{long2015fully}.
However, they do not discuss the topology nor the training parameters.
Furthermore, there is no performance comparison with network topologies in the literature used for semantic segmentation.

Here, we develop deep neural models specifically for the semantic segmentation task using state-of-the-art concepts and tools to train and predict seismic facies efficiently. Our goal is to reduce the amount of labeled data the geoscientist has to produce to have all the slices in one cube correctly segmented. In other words, with a minimum number of annotated images, we want to produce efficient and good quality segmentation in the rest of the cube.

We explore the network topology to reduce the number of parameters and operations but yet improve the performance of test data.
Additionally, we analyze the performance impact of different input sizes and number of training examples.
Finally, we compare our results to those obtained by using well-known topologies applied to the seismic facies segmentation task in the literature.

We divide this paper into five sections.
In the following section, we present the dataset and the pre-processing procedure.
Next, in the Network Architectures section, we present the techniques used to design our models and discuss the network topologies.
In the Experiments section, we describe training parameters and the experimental results.
Finally, we present the conclusions and future work in the last section.

\section{Dataset} \label{dataset}

In this section, we describe the dataset used in our experiments and the data preparation process.

\subsection{Penobscot Seismic Dataset}

The generation of seismic images is a sophisticated process which comprises many steps.
The first one is data acquisition, which requires an intense sound source, such as an air gun, to direct sound waves into the ground.
These waves pass through different layers of rock (strata) and are reflected, returning to the surface, where geophones or hydrophones record them.
This signal is then processed in an iterative procedure to generate the seismic images.
Finally, interpreters analyze the resulting images and discriminate the different categories -- or \textit{facies} \cite{mattos2017assessing}.
These categories represent the overall seismic characteristics of a rock unit that reflect its origin, differentiating this unit from the other ones around it \cite{mattos2017assessing}.

\begin{figure}[ht]
\centering
\includegraphics[width=.65\linewidth]{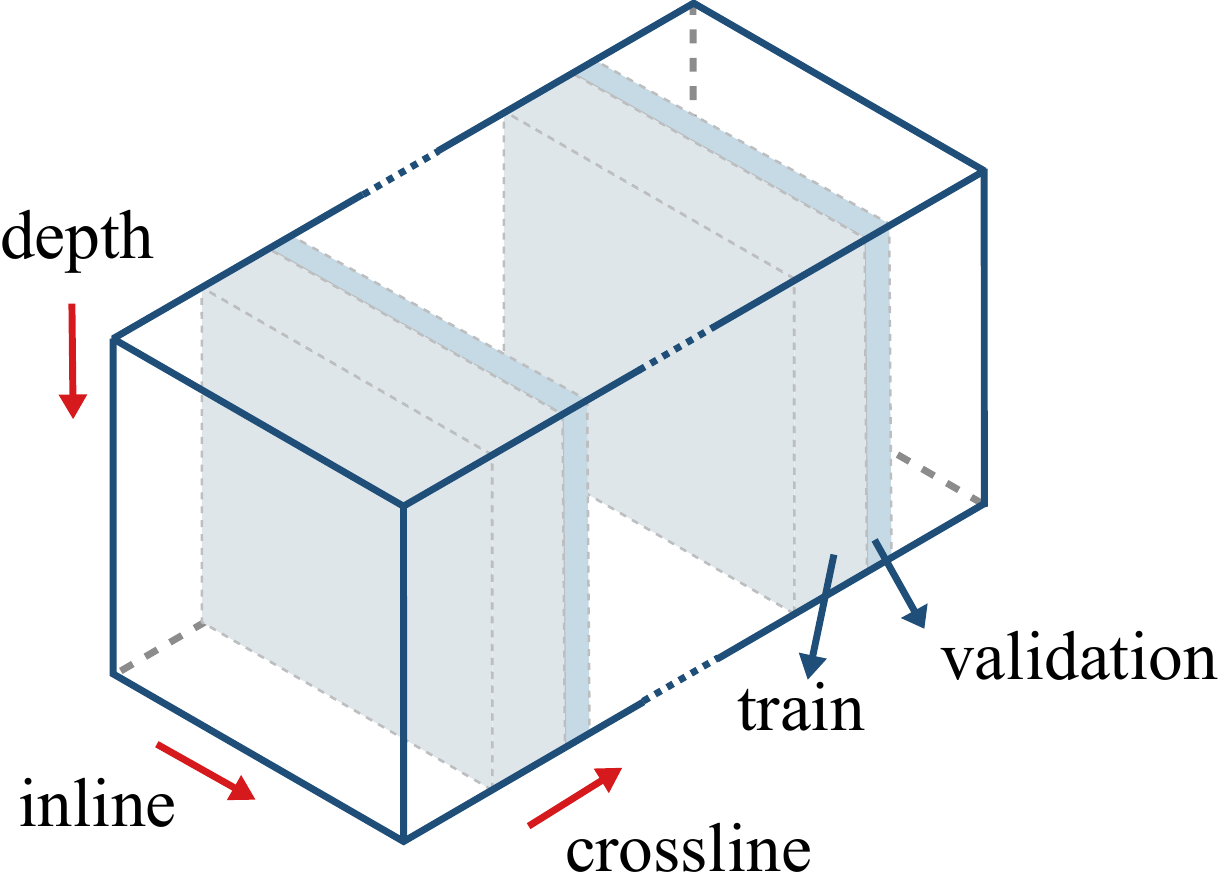}
\caption{3D seismic volume with the division highlighted in blocks of inline slices.
The red arrows indicate inline, crossline and depth directions.
In each block, the first slices go to the training set and the rest to the validation set.}
\label{fig:seismic_cube}
\end{figure}

Penobscot 3D Survey \cite{baroni2018penobscot} is the publicly available dataset used in this work.
It consists of a horizontal stack of 2D seismic images (\textit{slices}), creating a 3D volume as depicted in Figure~\ref{fig:seismic_cube}.
The vertical axis of this volume represents depth.
The remaining axes define the \textit{inline} and \textit{crossline} directions -- red arrows in Figure~\ref{fig:seismic_cube}. The inline slices are the images in the cube perpendicular to the inline direction, indicated by the blue blocks in Figure~\ref{fig:seismic_cube}. The same idea applies to the crossline slices, which are images along the depth axis and perpendicular to the crossline axis.
The Penobscot dataset contains 481 crossline and 601 inline slices, with dimensions $601 \times 1501$ and $481 \times 1501$ pixels, respectively.
It is important to mention that we used only inline slices in this work and we removed corrupted or poor-quality images, as indicated by an expert.
After removing these images, we ended up with 459 inline slices.

\begin{figure}[ht]
\centering
\includegraphics[width=.95\linewidth]{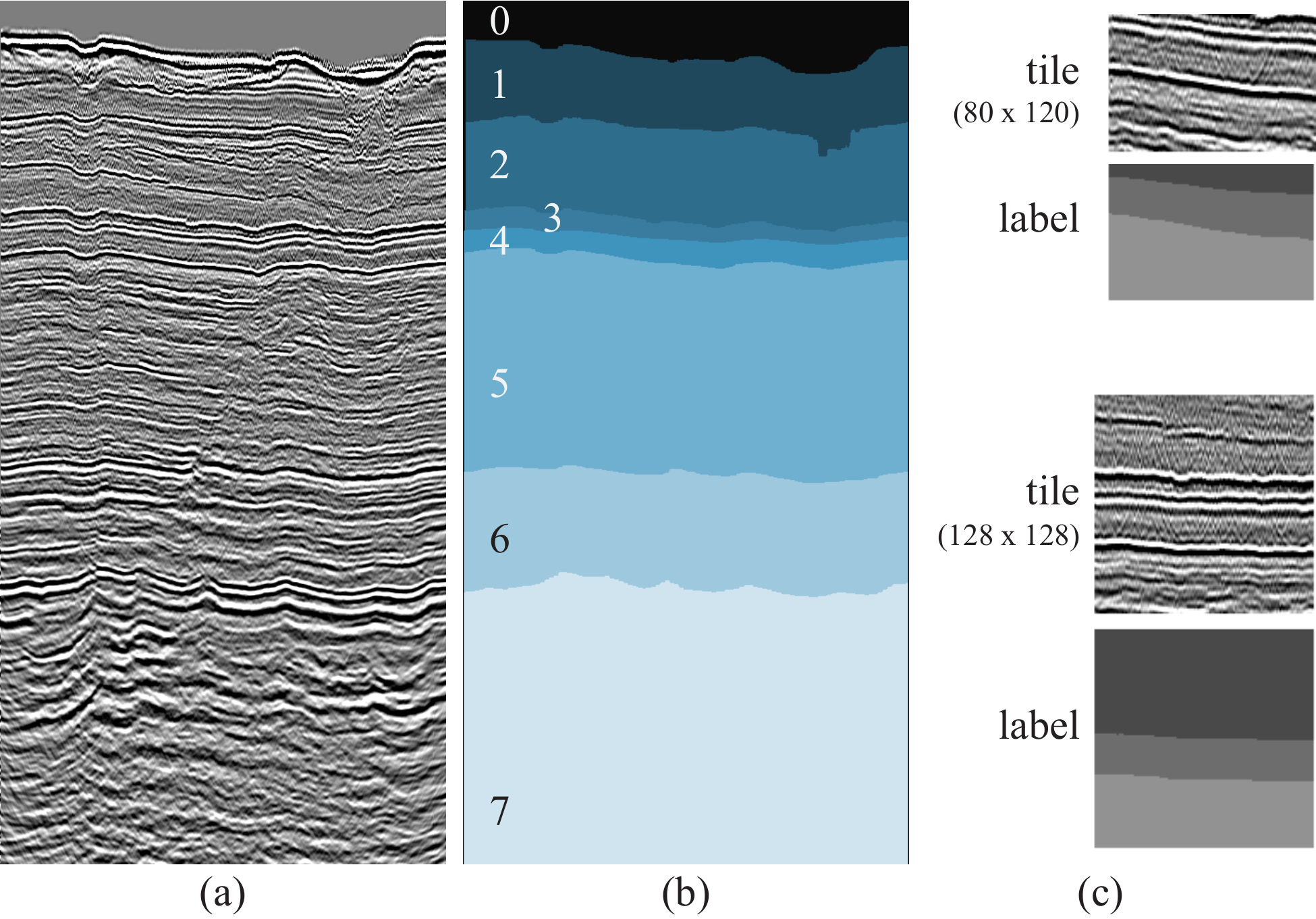}
\caption{(a) Example of an inline slice (we cropped the bottom of the image).
(b) The corresponding mask for the input image, showing the eight categories.
(c) Tile examples with their respective label masks.
For better visualization, the tiles are not on the same scale as (a) and (b).}
\label{fig:penob_example}
\end{figure}

Geoscientists interpreted and annotated the slices generating a label mask for each one.
The experts interpreted seven horizons: H1, H2, H3, H4, H5, H6, and H7, numbered from the lowest depth to the highest. They divide the seismic cube into eight intervals with different pattern configurations. Figure~\ref{fig:penob_example} b shows the 7 interpreted horizons, creating 8 classes.
The geoscientists based their interpretation on configuration patterns that indicate geological factors like lithology, stratification, depositional systems, etc. \cite{brown1980seismic}. In the following list we explain briefly the seismic facies of each of the horizon intervals based on the amplitude and continuity of reflectors.

\begin{enumerate}[start=0]
\item ocean. 
\item the facies unit is composed of high-amplitude reflectors. Although most of the reflectors are continuous, some have diving angles and others are truncated, evidencing a high energy environment.
\item the package consists mostly of parallel, high-amplitude reflectors.
\item reflectors are predominantly subparallel and present varying amplitude.
\item reflectors between these horizons are continuous but have low amplitude, which makes it difficult to identify them.
\item facies unit containing parallel to subparallel reflectors, like the previous interval, but less continuous.
\item the facies unit is characterized by parallel to subparallel, continuous, high-amplitude reflectors.
\item the facies unit below the 7$^\text{th}$ horizon is characterized primarily by parallel, concordant, high-amplitude reflectors. 
\end{enumerate}

\subsection{Dataset preparation}

As discussed in \cite{mattos2017assessing,chopra2006applications}, here we assume that one may identify different categories by their textural features.
Therefore, a model that can distinguish textures in an image can be used to separate seismic facies.

In our segmentation task, we chose to break the seismic images into smaller tiles (Figure~\ref{fig:penob_example} (c)), as the original ones are large and would require a more complex model to deal with such a big input size.
Also, the original images have a float range around [-30.000, 33.000] and we decided to rescale them to the grayscale range of [0, 255].

Our dataset preparation comprises the following steps:

\begin{enumerate}
  \item remove corrupted images (indicated by a domain expert);
  \item split image files into training, validation and test sets;
  \item process images removing outliers, and rescale values between 0-255;
  \item break processed images into tiles using a sliding window mechanism that allows overlap;
\end{enumerate}

As we want to simplify the interpreter job, we must create training, validation and test sets that correspond to the nature of this task, i.e., we should use only data from a single seismic cube. We acknowledge that adjacent slices can be similar since they represent seismic characteristics of neighboring regions. For the same reason, we also expect slices located far from each other in the volume to be less correlated. However, different cubes represent different regions that usually do not have interchangeable classes or characteristics. Then, for our application, we could not use data from other cubes to test overfitting.

Keeping that in mind, we used the following scheme to split the cube into training, validation and test sets (see Figure~\ref{fig:seismic_cube}):
\begin{enumerate}
  \item divide the cube into \textit{n} blocks;
  \item in each block, select the first 70\% slices for the training set and the remaining 30\% for the validation set;
  \item (optionally) limit the total number of training slices to be \textit{x} and randomly select $x$/$n$ slices in each block so we can simulate limited data scenarios;
\end{enumerate}

Following the work in \cite{baroni2018penobscot} we merged classes 2 and 3, as class 3 represents a very thin layer (about 20 pixels of depth).
Therefore, the final number of classes is seven.

\section{Network architectures} \label{network_arch}

In this paper, we present two new topologies built explicitly for the semantic segmentation of seismic data, Danet-FCN2 and Danet-FCN3.
As mentioned before, this kind of image can be very different from common texture images, such as the images in the DTD dataset mentioned in the introduction, where tiles representing textures are larger and have colors that make classes more separable.

The proposed architectures use residual blocks \cite{he2016deep} to extract features (encoder), and a novel transposed unit of the same structure to reconstruct the original image with its respective labels (decoder).
Contrarily to the previously developed Danet-FCN \cite{chevitarese2018seismic}, these newer models do not have additional fused connections between the encoder and the decoder.
In \cite{long2015fully}, the authors incorporated skip connections in the topology to combine coarse, high layer information with detailed, low layer information.
On the other hand, He and his colleagues show that shortcut connections and identity after-addition activation make information propagation smooth \cite{he2016deep}.
In this sense, we merge FCN and ResNet ideas by removing the shortcuts as in FCN, and U-Net, and by replacing the contracting path with residual units.
For the expansive part, however, we had to create the inverse operation that we called transposed residual since it was inexistent.
In addition to a gain in performance, our new architecture, composed of residual blocks, already implements shortcut paths along with the skip connection, what makes fuse (FCN and U-Net) operations unnecessary.

\begin{figure}[ht]
\centering
\includegraphics[width=.99\linewidth]{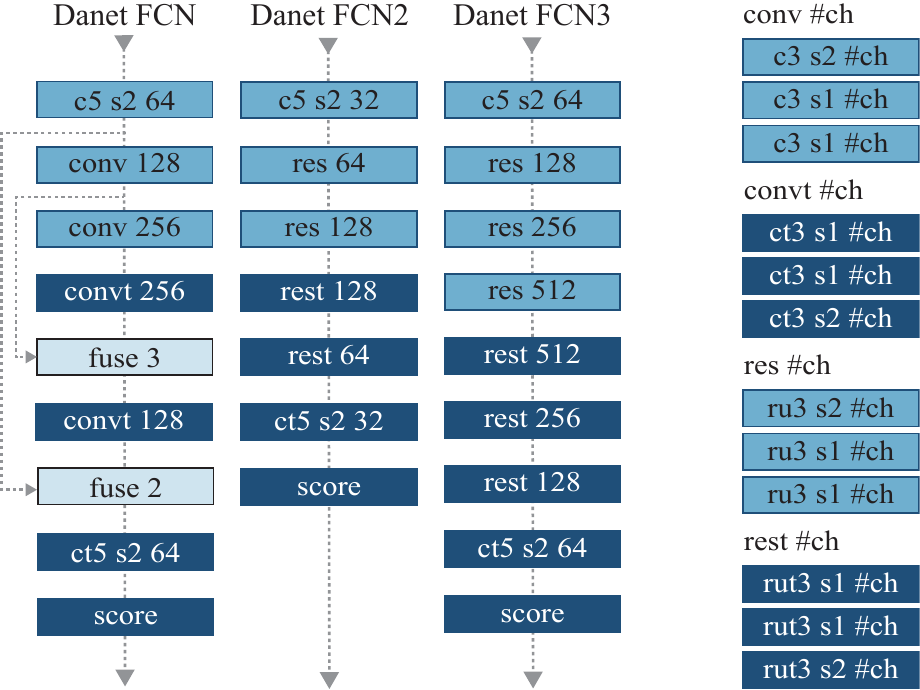}
\caption{These are the topologies of the three models.
On the right, we present the description of each block, where \#ch is the number of channels (filters) used in all operations inside it.}
\label{fig:topologies}
\end{figure}

In Figure~\ref{fig:topologies} we compare all three Danet-FCN topologies, where \textbf{c} stands for a convolution layer, \textbf{s} is the stride, and \textbf{ru} stands for a residual unit.
So, an entry $c5~s2~64$ stands for a convolution with $64$ filters of size $5 \times 5$ and stride $2$.
The residual configuration we use in this work is the same one presented in \cite{chevitarese2018efficient}, described by Equation~\ref{eq:resunit}:

\begin{equation}
\begin{aligned}
\label{eq:resunit}
\mathbf{x} &\leftarrow h ( \mathbf{x} ) + \mathcal{F}\left ( \mathbf{x}, \theta \right ) \\
\mathbf{y} &\leftarrow f( \mathbf{x} )
\end{aligned}
\end{equation}

\noindent
where $h(\mathbf{x})$ is an identity mapping, and $f(\mathbf{x})$ is a ReLU \cite{NIPS2012_4824} function.

\subsection{Danet-FCN}

\cite{chevitarese2018seismic} introduced this model that is based on the VGG-FCN topology presented by \cite{long2015fully}. In \cite{long2015fully}, the authors convert the fully connected layers at the end of the network structure into $1 \times 1$ convolutions, and append a transposed convolution to up-sample its input features to the input image size.
We decided to revisit this network not only to compare it with the ones proposed here but also to extend the discussion about it.
The main idea behind Danet-FCN is to have a simpler network -- instead of using VGG-size structures -- that leads to faster training times, and less memory usage.
In contrast to its relative, VGG-FCN, Danet-FCN only performs convolutions, and strides in its operations replace pooling to reduce data dimensionality.
We believe that this is one of the reasons that made Danet-FCN perform remarkably well on the datasets in which it was tested.

\subsection{Danet-FCN2}

\begin{figure}[ht]
\centering
\includegraphics[width=.60\linewidth]{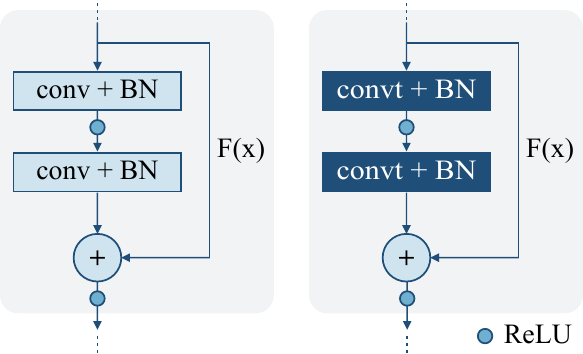}
\caption{On the left we have the regular residual structure.  On the right, we have the proposed transposed residual unit.}
\label{fig:residual}
\end{figure}

In this work, we propose the use of transposed convolutions to create transposed residual units producing a ``mirror'' structure similar to the one in U-Net \cite{unet}.
Figure~\ref{fig:residual}
Although this operation can be more expensive comparing to dilated convolutions, we noticed that it can produce smoother results.
The fact that the upscale is learned instead of being predefined by the user may be one explanation for that.
Another nice feature we obtained with this new residual structure is the creation of links between the encode layers and decode ones, similar to the fuse we have in FCN-16 and FCN-8.
Those links improve the segmentation results substantially, as discussed in \cite{long2015fully}.

Figure~\ref{fig:topologies} shows the entire structure of Danet-FCN2.
It has a very compact topology allowing it to learn from a minimal number of examples -- see Table~\ref{tb:results}.
Also, we noticed that this model is the fastest to train in most of the configurations too.

\subsubsection{Danet-FCN3}

Our biggest topology is similar to Danet-FCN2, but comprising more residual units with more filters.
Comparing to U-Net it has more parameters on one hand, but less operations to compute on the other hand.
It is expected a deterioration of the results for smaller datasets since Danet-FCN3 has much more parameters compared to its relatives Danet-FCN and Danet-FCN2.

\section{Experiments} \label{experiments}

Following the idea presented in \cite{chevitarese2018efficient}, we want to investigate whether the proposed topologies are useful in a limited data context, which is similar to analyze the influence of the number of slices available for training in the models' performance.
Also, there are only a few small datasets publicly available for training in the Oil and Gas industry.
It is important to highlight that the process of annotating seismic images is expensive and time-consuming.
Hence, finding the minimum number of labeled images required to train a model successfully, is valuable for the O\&G industry.
In this section, we describe the experiments we conducted to accomplish our goals.

We generated different tile datasets from the seismic images by varying the \textit{number of slices} for the training set: 100, 13, 9 and 5 slices.
This allows us to simulate the limited dataset scenarios and analyze models' performance with a decreasing number of examples for training.
Based on the results presented in \cite{chevitarese2018seismic}, we selected a tile size of  $80 \times 120$ pixels.
In addition, we generated tiles of $128 \times 128$ pixels so we could compare our networks with relevant models in the literature of semantic segmentation, such as U-Net \cite{unet}, which expects larger input images.
For both tile sizes, we let 50\% of overlap in each direction.
This means that the sliding window skips, for example, $60$ pixels horizontally and $40$ pixels vertically in the $80 \times 120$ tile case.
Table \ref{tb-dataset} summarizes all the generated tile datasets and shows the number of training examples in each one.

We compare our models with slightly modified networks from the literature: FCN16, FCN8 \cite{long2015fully} and U-Net \cite{unet}.
In their original applications, the models were designed to handle input images larger than our tiles, but considerably smaller than an entire seismic image.
Therefore, to fairly compare our models with FCN and U-Net, we decided to make minor changes in their architecture to accommodate the smaller or non-square tiles.
For the FCN networks, we added a crop mechanism to correct shapes of feature maps in the upsampling part.
For the selected tile sizes, we cropped at most 1 pixel from the feature maps, which we can consider that will not affect the results.
For the U-Net, we changed the unpadded convolutions to padded convolutions, keeping intact the rest of the structure.

It is important to mention that, from the selected models in the literature, we only trained FCN-8 using both tile sizes.
The other ones would require further changes, or cropping a significant number of pixels in order to comply with the input size of $80 \times 120$.

\subsection{Training parameters}

In all experiments, we fixed the mini-batch as $64$, the weight decay coefficient as $5 \times 10^{-4}$ and the maximum number of epochs as $200$.
We also used the RMSProp optimization algorithm \cite{rmsprop} with $\nu = 0.9$, $\varphi = 0.9$ and $\varepsilon = 1.0$ in all training sessions.
The Xavier method \cite{glorotInit} of initialization was used in all convolutional kernels while the biases were initialized with zeros.

The Danet models have batch normalization layers and we chose their parameters $\epsilon = 1 \times 10^{-5}$ and $\mu = 0.997$.
We also used a learning rate (lr) schedule for the Danet models, to take advantage of the batch normalization benefits.
Starting with a relatively high lr of $0.01$, we performed a stepwise decrease:
\begin{itemize}
\item epoch = $50$, lr = $0.001$
\item epoch = $100$, lr = $5 \times 10^{-4}$
\item epoch = $150$, lr = $1 \times 10^{-5}$
\end{itemize}

For the FCN and U-Net models, we used a constant learning rate of $1 \times 10^{-4}$, and dropout with $\rho = 0.5$, as indicated in the original papers.
Also, for the FCN models, we used the same scheme of loading the weights of a  VGG16 network previously trained with the ImageNet dataset \cite{Imagenet}.

We trained the models using Tensorflow 1.9, 4 or 2 GPUs K80 (depending on the number of examples) in a Power8 node.
Also, we adopted the mean intersection over union (mIOU) in the validation dataset as the metric to select the best model during training.
One should notice that we calculated the mIOU by averaging the IOU over all classes, with the same weights, even though they are not balanced.


\begin{table}[ht]
\caption{Datasets generated for the experiments}
\label{tb-dataset}
\centering
\begin{tabular}{@{}clcccc@{}}
\toprule
\multirow{2}{*}{$80 \times 120$} & Images & 100 & 13 & 9 & 5 \\ \cmidrule(l){2-6} 
 & Tiles & 15392 & 2000 & 1376 & 768 \\ \midrule
\multirow{2}{*}{$128 \times 128$} & Images & 100 & 13 & 9 & 5 \\ \cmidrule(l){2-6} 
 & Tiles & 7792 & 1008 & 688 & 384 \\ \bottomrule
\end{tabular}
\end{table}

\section{Results and discussion} \label{results}

To compare the trained models, we used the 40 images selected for the test set, as mentioned before.
Our test procedure for all trained models consists of the following steps:

\begin{enumerate}
  \item For each image, we break it into non-overlapping tiles of the same size the model was trained with;
  \item we predict the tile masks with the trained model;
  \item the entire mask for that image is assembled by uniting the predicted mask tiles;
  \item once the entire predicted mask is formed, we calculate the IOU for each class in that image and average the results to get the mIOU for the image;
  \item finally, when we have the mIOU for all 40 images, we average the results to get the overall response on the test set, which we named here mmIOU.
\end{enumerate}

\begin{figure*}[ht]
\centering
\includegraphics[width=.9\linewidth]{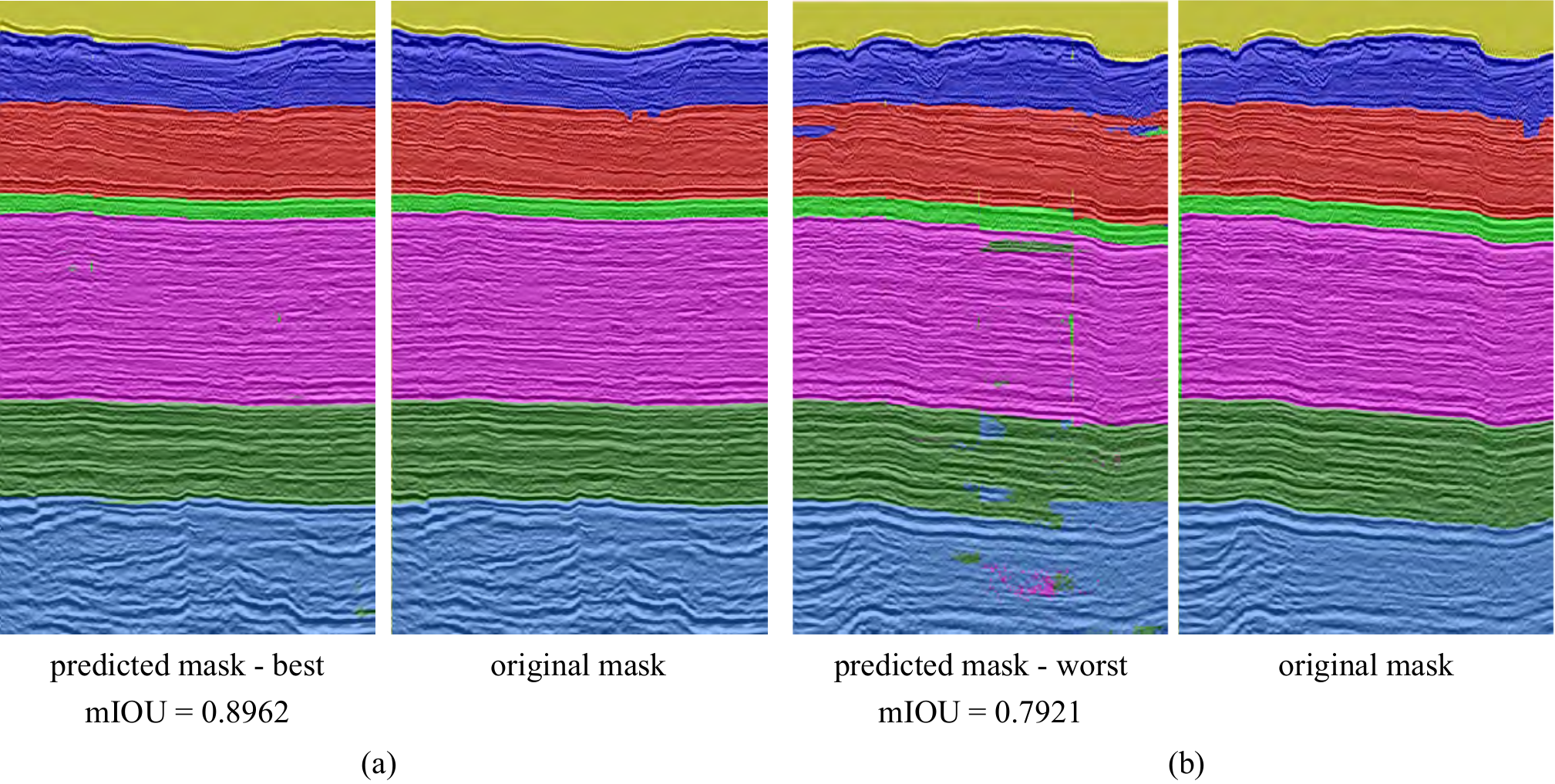}
\caption{Masks generated by Danet-FCN3 on the test set, for tiles of size 80 x 120.
(a) Best results for the 5 slices case.
(b) Worst results for the 5 slices case.}
\label{fig:resultsD3}
\end{figure*}

The test procedure enables us to compare the models both quantitatively (mmIOU) and qualitatively (predicted masks).
Table \ref{tb:results} summarizes the mmIOU for all the experiments.

\begin{table}[ht]
\caption{Experiments results.}
\label{tb:results}
\resizebox{.99\linewidth}{!}{
\begin{tabular}{@{}cccccc@{}}
\toprule
\multirow{2}{*}{Image size} & \multirow{2}{*}{Model} & \multicolumn{4}{c}{Number of slices (mmIOU)} \\
 &  & 5 & 9 & 13 & 100 \\ \midrule
\multirow{4}{*}{$80 \times 120$} & DanetFCN & \textbf{0.9180} & 0.9592 & 0.9700 & 0.9877 \\
 & DanetFCN2 & 0.8932 & 0.9609 & 0.9616 & 0.9874 \\
 & DanetFCN3 & 0.8385 & \textbf{0.9689} & \textbf{0.9764} & \textbf{0.9900} \\
 & FCN8 & 0.5819 & 0.7196 & 0.7389 & 0.9364 \\ \midrule
\multirow{6}{*}{$128 \times 128$} & DanetFCN & \textbf{0.8150} & 0.8366 & 0.8272 & 0.9413 \\
 & DanetFCN2 & 0.7618 & \textbf{0.8800} & 0.8513 & 0.9360 \\
 & DanetFCN3 & 0.7989 & 0.8484 & \textbf{0.8622} & \textbf{0.9485} \\
 & FCN8 & 0.4497 & 0.5790 & 0.6605 & 0.9057 \\
 & FCN16 & 0.1524 & 0.4997 & 0.5950 & 0.9369 \\
 & U-Net & 0.1397 & 0.1431 & 0.1371 & 0.1880 \\ \bottomrule
\end{tabular}}
\end{table}

For the tile size of $80 \times 120$ pixels, all Danet models presented a mmIOU higher than 80\%, even in the very limited dataset of 5 slices.
One can notice that the results deteriorate with the decrease in the number of examples, but this effect is stronger in the FCN8 and FCN16.
In the non-limited context (100 slices in the training set), both FCN8 and FCN16 have a performance similar to the Danet models, but Danet-FCN3 beats them for both tile sizes.
On the other hand, we were not able to effectively train the U-Net model with our datasets, as shown in the last line of Table \ref{tb:results}.
The training loss did not converge and the results were poor.
Further investigation is required to understand the reason for that.
A first assumption would be that the input size is still too small for this network.

The qualitative results for Danet-FCN3 can be seen in Figure \ref{fig:resultsD3} and Figure \ref{fig:resultsD3}.
In the 100 slices case, the predicted mask is almost identical to the original one and, in the limited data context, the result is still very close to the mask generated by a human expert.
Table \ref{tb:iou_classes} complements Figures \ref{fig:resultsD3} and \ref{fig:resultsD3} as it shows the IOU for each class separately.
From these values, we can see that the classes most affected by the reduced amount of examples are classes 1, 3 and 5, but the other ones still maintain an IOU higher than 80\%.
The predicted masks along with the high values of mIOU are a strong indication that our topologies are well suited for the semantic segmentation of seismic images even with a restricted number of training examples available.

\begin{table}[ht]
\caption{IOU per class Danet-FCN3 and tiles $80 \times 120$.}
\label{tb:iou_classes}
\resizebox{.99\linewidth}{!}{
\begin{tabular}{@{}cccccccccc@{}}
\toprule
 &  & \multicolumn{8}{c}{IOU per class} \\ \cmidrule(l){3-10} 
slices & case & 0 & 1 & 2 & 3 & 4 & 5 & \multicolumn{1}{c|}{6} & mIOU \\ \cmidrule(l){3-10} 
5 & best & 0.888 & 0.531 & 0.983 & 0.887 & 0.997 & 0.989 & \multicolumn{1}{c|}{0.998} & 0.896 \\
 & worst & 0.889 & 0.350 & 0.897 & 0.596 & 0.968 & 0.855 & \multicolumn{1}{c|}{0.989} & 0.792 \\
100 & best & 0.991 & 0.991 & 0.999 & 0.990 & 1.000 & 0.998 & \multicolumn{1}{c|}{1.000} & 0.995 \\
 & worst & 0.970 & 0.983 & 0.994 & 0.981 & 0.999 & 0.995 & \multicolumn{1}{c|}{0.995} & 0.988 \\ \bottomrule
\end{tabular}}
\end{table}

\begin{table}[ht]
\caption{Computational resources.}
\label{tb:profile}
\resizebox{.99\linewidth}{!}{
\begin{tabular}{@{}ccccccc@{}}
\toprule
Model name & DanetFCN & DanetFCN2 & DanetFCN3 & FCN8 & FCN16 & UNET \\ \midrule
\begin{tabular}[c]{@{}l@{}}Millions of \\ parameters\end{tabular} & 4.46 & 6.66 & 39.2 & 134.31 & 134.34 & 31.03 \\ \midrule
\begin{tabular}[c]{@{}l@{}}Millions of \\ operations\end{tabular} & \multirow{2}{*}{3213} & \multirow{2}{*}{1880} & \multirow{2}{*}{10572} & \multirow{2}{*}{9032} & \multirow{2}{*}{-} & \multirow{2}{*}{-} \\
tile = 80 x 120 &  &  &  &  &  &  \\ \midrule
\begin{tabular}[c]{@{}l@{}}Millions of \\ operations\end{tabular} & \multirow{2}{*}{5477} & \multirow{2}{*}{3140} & \multirow{2}{*}{17675} & \multirow{2}{*}{14094} & \multirow{2}{*}{14093} & \multirow{2}{*}{24133} \\
tile = 128 x 128 &  &  &  &  &  &  \\ \bottomrule
\end{tabular}}
\end{table}

Table \ref{tb:profile} compares the models by the number of parameters and operations.
Danet-FCN has the lowest number of parameters and, at the same time, is the best performing model for the 5 slices training case for both tile sizes.
Danet-FCN3 and U-Net have parameters in the same order of magnitude, but U-net has significantly more operations.
In this case, we were able to take advantage of the deep residual network and get excellent results while keeping the number of operations similar to what we find in FCN and U-Net.

Comparing Table \ref{tb:profile} with Table \ref{tb:results} we can conclude that Danet-FCN2 can be seen as the model with the best balance between performance and efficiency.
It has the lowest number of operations and almost 5 times less parameters than U-net while presenting mmIOU above 89\% for all cases with $80 \times 120$ tiles.

\section{Conclusion} \label{conclusion}

In this work, we designed and compared deep neural models specifically for the task of semantic segmentation of seismic images, which have different characteristics than standard images.
Our proposed networks combine different features from existing architectures and a novel feature which is the use of a transposed convolution to create transposed residual units producing a "mirror structure".
To the best of our knowledge this feature is novel and could be tested as part of neural models for other types of images and even for other tasks.

Our experiments show that our proposed models lead to significantly better results than the semantic segmentation neural models designed for standard images.
Furthermore, we evaluated these models under the realistic condition that there would be very few examples for training (as few as 5 slices) and showed that we can still achieve high values in the intersection over union metric and, through visual inspection, see a good reproduction of the original masks.

For future work we want to test the proposed topologies on new seismic datasets.
Also, it is important to experiment with more seismic layers being considered by the model, and then increasing the number of facies (classes) recognized by the network.
Finally, we want to include to the current data (texture only) the geophysical properties (e.g., velocity).

\bibliographystyle{named}
\bibliography{ijcai19-reduced}

\end{document}